\documentclass[a4paper,11pt]{article}
\usepackage{jcappub} % for details on the use of the package, please
\usepackage{color}
\usepackage{xcolor}
\usepackage[T1]{fontenc} % if needed

\newcommand{\red}[1]{{\color{red}{#1}}}

\title{\boldmath Strong field gravitational lensing by a charged Galileon black hole}

\author{Shan-Shan Zhao}
\author[1]{and Yi Xie\note{Corresponding author.}}

\affiliation{School of Astronomy and Space Science, Nanjing University, Nanjing 210093, China\\
Shanghai Key Laboratory of Space Navigation and Position Techniques, Shanghai 200030, China\\
Key Laboratory of Modern Astronomy and Astrophysics, Nanjing University, Ministry of Education, Nanjing 210093, China}

\emailAdd{clefairy035@163.com}
\emailAdd{yixie@nju.edu.cn}

\abstract{Strong field gravitational lensings are dramatically disparate from those in the weak field by representing relativistic images due to light winds one to infinity loops around a lens before escaping. We study such a lensing caused by a charged Galileon black hole, which is expected to have possibility to evade no-hair theorem. We calculate the angular separations and time delays between different relativistic images of the charged Galileon black hole. All these observables can potentially be used to discriminate a charged Galileon black hole from others. We estimate the magnitudes of these observables for the closest supermassive black hole Sgr A*. The strong field lensing observables of the charged Galileon black hole can be close to those of \red{a tidal Reissner-Nordstr\"{o}m black hole or those of a Reissner-Nordstr\"{o}m black hole}. It will be helpful to distinguish these black holes if we can separate the outermost relativistic images and determine their angular separation, brightness difference and time delay, although it requires techniques beyond the current limit.
}


\begin{document}
\maketitle
\flushbottom

\section{Introduction}
\label{sec:intro}

Strong field gravitational lensings have received much attention in recent years because they are dramatically disparate from those in the weak field both on mathematical descriptions and on astronomical observations. It was firstly illuminated by Darwin in 1959 \cite{Darwin1959PRSLSA249.180} that light bending by a compact body can exceed $2\pi$ and the light even can wind several loops before escaping, which develops infinite discrete images on two sides of the body closely, called relativistic images. Relativistic images, not predicted by the classical weak gravitational field lensing, provide a new way to study the properties of spacetime in the strong gravitational field. Plenty works have done on the strong field lensings by a Schwarzchild black hole \cite{Virbhadra2000PRD62.084003,Bozza2001GRG33.1535,Eiroa2004PRD69.063004,KantiDey2012arXiv1208.3306}, by static and spherically symmetric spacetimes \cite{Bozza2002PRD66.103001,Perlick2004PRD69.064017,Amore2007PRD75.083005,Raffaelli2014arXiv1412.7333}, by a Kerr black hole \cite{Bozza2003PRD67.103006,Bozza2005PRD72.083003,Bozza2006PRD74.063001,Bozza2008PRD78.063014} and by a Kerr black hole in the presence of the cosmological constant \cite{Kraniotis2011CQG28.085021} and electrically charge \cite{Kraniotis2014GRG46.1818}.

With rapid development of advanced technology, the strong filed gravitational lensings have become appealing observational effects for astronomy and fundamental physics. Relativistic images might be helpful for making a better understanding of different black holes \cite{Bozza2002PRD66.103001,Bin-Nun2010PRD81.123011,Gyulchev2007PRD75.023006,Gyulchev2013PRD87.063005,Herdeiro2014PRL112.221101,Cunha2015PRL115.211102} and might be able to provide observational evidences for the possible existence of naked singularities \cite{Virbhadra2002PRD65.103004,Virbhadra2008PRD77.124014,Sahu2012PRD86.063010,Sahu2013PRD88.103002,Gyulchev2008PRD78.083004} as well as wormholes \cite{Kuhfittig2014EPJC74.2818,Kuhfittig2015arXiv1501.06085,Nandi2006PRD74.024020,Tsukamoto2012PRD86.104062}. It also provides a promising way to test fundamental theories of gravitation in the strong field \cite{Virbhadra1998AA337.1,Eiroa2014EPJC74.3171,Sotani2015PRD92.044052}. However, it is still a challenge to observe such effects under the current observational capabilities. For the most possible candidates, the supermassive black holes in our galaxy Sagittarius A* (Sgr A*) and in M87, the angular separations between the relativistic images and the center of the lens are respectively tens micro-arcsecond ($\mu$as) and few $\mu$as. The best ability for observing Sgr A* is at the level of $30$ $\mu$as at present \cite{Broderick2014ApJ784.7}, not sufficient for detection. Besides their positions, the brightness of these images are other observables. Under the assumption that the relativistic images can be recognized, several works have studied the light curves of the stars moving around Sgr A* \cite{Bozza2004ApJ611.1045,Bin-Nun2010PRD82.064009,Bin-Nun2011CQG28.114003}, and the S14 is proved to be the best candidate \cite{Bozza2005ApJ627.790}. If the light source is a pulsar or a celestial body with time signals, time delays among different relativistic images are observables as well \cite{Bozza2004GRG36.435,Virbhadra2008PRD77.124014,Man2014JCAP11.025}. A review of gravitational lensing by black holes in the strong field can be found in ref. \cite{Bozza2010GRG42.2269}.

One difficulty in describing the strong filed gravitational lensing is that we cannot use the methodology of small deflection angle approximation which works very well in the weak field. It has been proved that when a photon moves around a black hole, there exists an innermost unstable orbit named photon sphere \cite{Atkinson1965AJ70.517,Claudel2001JMP42.818}. The deflection angle will diverge when a photon approaches the photon sphere. In order to dispose this divergence in a static, spherically symmetric and asymptotically flat spacetime, one way is to expand the function of the deflection angle near the photon sphere and obtain an approximate analytical solution \cite{Bozza2001GRG33.1535,Bozza2002PRD66.103001}. The method can give the deflection angle in the strong deflection limit (SDL) \cite{Bozza2010GRG42.2269}. A logarithmic approximation is used to solve the deflection angle integral, which could make the formula conciser in presentation and easier to handle than other kinds of approximations \cite{Darwin1959PRSLSA249.180,Amore2006PRD73.083004,Amore2006PRD74.083004,Amore2007PRD75.083005,Iyer2007GRG39.1563}. A lens equation is also needed to define the geometrical relationships among the observer, the lens, the source and the images.

Thus, if a specific spacetime and a lens equation are both known, the observables, including the positions, magnifications and time delays of relativistic images, can be analytically deduced in the SDL, whose coefficients may vary in alternative theories of gravity. Although observations and experiments have proven the validation of Einstein's general relativity (GR) \cite{Will1993TEGP,Will2006LRR9.3}, it seems that the theory might be incomplete. It is difficult for GR to explain the flat rotation curves of spiral galaxies \cite{Rubin1970ApJ159.379,Roberts1975ApJ201.327,Sofue2001ARAA39.137} without introducing dark matter and the present acceleration of the Universe \cite{Riess1998AJ116.1009,Perlmutter1999ApJ517.565} without dark energy. Nevertheless, the physical nature of dark matter and dark energy remains still unknown. Another way to solve the problems is to modify the theory of gravity and these modified theories can generate interesting astrophysical and cosmological consequences \cite{Clifton2012PhR513.1}. Strong field gravitational lensings can provide a possible way to test and distinguish theoretical predictions in the vicinity of a compact body by these modified theories of gravity and GR \cite{Bhadra2003PRD67.103009,Eiroa2005BrJPh35.1113,Eiroa2006PRD73.043002,Psaltis2008LRR11.9,Eiroa2011CQG28.085008,Eiroa2012PRD86.083009,Eiroa2013PRD88.103007,Chen2009PRD80.024036,Chen2015JCAP10.002,Ding2011PRD83.084005,Ding2013PRD88.104007,Wei2011PRD84.041501,Wei2012JCAP10.053,Wei2015AHEP2015.454217,Wei2015EPJC75.253,Wei2015EPJC75.331,Deng2012IJTP51.1632,Sadeghi2014JCAP06.028,Sarkar2006CQG23.6101}. In this work, we will study the string field lensing in the SDL by a charged Galileon black hole.

Galileon model was recently proposed as a scalar field theory \cite{Nicolis2009PRD79.064036,Deffayet2009PRD79.084003,Deffayet2009PRD80.064015,Deffayet2011PRD84.064039,Deffayet2013CQG30.214006} and its Lagrangian contains second-order derivatives of the scalar field, which leads to equations of motion of second order. It also turns out to be equivalent to the Horndeski theory in the 1970s \cite{Horndeski1974IJTP10.363}. On the largest scale, Galileon can explain the present cosmological acceleration without introducing dark energy, which make the theory attractive in cosmology \cite{Chow2009PRD80.024037,Felice2010PRL105.111301,Burrage2011JCAP1.014}. Meanwhile, Galileon can hide its effects from the Solar System tests by the Vainshtein screening mechanism \cite{Vainshtein1972PLB39.393,Babichev2010PRD82.104008,Babichev2013CQG30.184001} so that the parameterized post-Newtonian limit of Galileon is consistent with the one of GR \cite{Hohmann2015PRD92.064019}. It makes Galileon black holes \cite{Rinaldi2012PRD86.084048,Cisterna2014PRD89.084038,Sotiriou2014PRL112.251102,Sotiriou2014PRD90.124063} become important testbeds due to their strong gravitational fields. In fact, a black hole in the Galileon field is expected to have possibilities to evade the no-hair theorem and maintain non-trivial hairs \cite{Sotiriou2014PRL112.251102,Sotiriou2014PRD90.124063}, because the time information in the time dependent scalar field could be saved under the Galileon shift symmetry, although the hairs are perhaps unstable for static black holes \cite{Hui2013PRL110.241104} or under perturbations \cite{Ogawa2015arXiv1510.07400}. Black hole solutions with a time-dependent Galileon have been found \cite{Babichev2014JHEP08.106} and charged Galileon black holes solutions coupled with an Abelian gauge field have also been worked out \cite{Babichev2015JCAP5.31}. It was pointed out \cite{Hui2012PRL109.051304} that a positional offset between the stellar center and the centric black hole in a galaxy in Galileon theory might exist for detection, while strong field gravitational lensing by a Galileon black hole may provide another way for testing this theory.

In this work we will study the strong field lensing by a charged Galileon black hole \cite{Babichev2015JCAP5.31} in the SDL \cite{Bozza2002PRD66.103001,Bozza2004GRG36.435}. \red{Although it is unlikely to find an astrophysical charged black hole since any electric charge would be easily neutralized, this work will theoretically relate the charged Galileon black hole to its observational properties, which might be helpful to understand effects of the Galileon model in astrophysical observable quantities.} In section~\ref{sec2}, the spacetime of the charged Galileon black hole is briefly reviewed for completeness. The lens equation with asymptotically flat approximation and the assumption of the source location  are dicussed in section~\ref{sec3}. The strong field lensing by the charged Galileon black hole in the SDL is calculated in section~\ref{sec4}. The observables of such a lensing, including the positions, the brightness and time delays between relativistic images can be found in section~\ref{sec5}. Some estimations of observables for Sgr A*  and comparisons of the results among the charged Galileon black hole, a Reissner-Nordstr\"{o}m (RN) black hole and \red{a tidal RN black hole \cite{Dadhich2000PLB487.1} which is a braneworld black hole} are also been given. Finally, in section~\ref{sec6}, we summarize our results and discuss their implication.

%%%%%%%%%%%%%%%%%%%%%%%%%%%%%%%%%%%%%%
%%%%%SECTION2%%%%%%%%%%%%%%%%%%%%%%%%%%%
%%%%%%%%%%%%%%%%%%%%%%%%%%%%%%%%%%%%%%

\section{Spacetime of a charged Galileon black hole}

\label{sec2}

The spacetime of the charged Galileon black hole will be briefly reviewed for completeness in this section, which only covers necessary information for our following work. More details about charged Galileon black holes can be found in ref. \cite{Babichev2015JCAP5.31}. We consider a Galileon action as \cite{Babichev2015JCAP5.31}
\begin{eqnarray}
  \label{eq:action}
  S[g_{\mu\nu},\phi,A_{\mu}] & = & \int\sqrt{-g}\mathrm{d}^4x\bigg[R-2\Lambda-\frac{1}{4}F_{\mu\nu}F^{\mu\nu} +\epsilon G_{\mu\nu}\nabla^{\mu}\phi\nabla^{\nu}\phi-\eta(\partial\phi)^2-\gamma T^{(M)}_{\mu\nu}\nabla^{\mu}\phi\nabla^{\nu}\phi\bigg],\nonumber\\
\end{eqnarray}
where the action $S$ is a function of the metric tensor $g_{\mu\nu}$, the scalar field $\phi$ and the magnetic potential of the standard Maxwell gauge field $A_{\mu}$; $g=\mathrm{det}(g_{\mu\nu})$ is the determinant of the metric tensor $g_{\mu\nu}$; $R$ is the Ricci scalar; $\Lambda$ is the cosmology constant; $F_{\mu\nu}\equiv\partial_{\mu}A_{\nu}-\partial_{\nu}A_{\mu}$ is the covariant tensor of the gauge field strength and $F^{\mu\nu}$ is the contravariant term; $G_{\mu\nu}$ is the Einstein tensor; and $T^{(M)}_{\mu\nu}$ is the energy-momentum tensor of the Maxwell field and is defined as
\begin{equation}
\label{}
T^{(M)}_{\mu\nu}\equiv \frac{1}{2}\bigg(F_{\mu\sigma}F^{\phantom{\nu}\sigma}_{\nu}-\frac{1}{4}g_{\mu\nu}F_{\alpha\beta}F^{\alpha\beta}\bigg).
\end{equation}
$\nabla$ is the nabla operator; $\epsilon$, $\gamma$ and $\eta$ are constants. $\epsilon$ indicates the non-minimal kinetic coupling between the scalar field and the gravity, $\gamma$ is a coupling constant of the gauge field to the scalar field, and $\eta$ represents the self-coupling of the scalar field. The case of $\epsilon=0$, $\gamma=0$ and $\eta=1/2$ corresponds to the minimally coupled situation.

In order to obtain solutions of black holes, some assumptions are needed and they are \cite{Babichev2015JCAP5.31}
\begin{enumerate}
	\item the metric tensor $g_{\mu\nu}$ is static and spherically symmetric as
\begin{equation}
  \label{eq:metric}
  \mathrm{d}s^2=-h(r)\mathrm{d}t^2+\frac{\mathrm{d}r^2}{f(r)}+r^2(\mathrm{d}\theta^2+\sin^2{\theta}\mathrm{d}\varphi^2);
\end{equation}
  \item the scalar field $\phi$ is linearly time-dependent as
\begin{equation}
\phi(t,r)=qt+\psi(r);
\end{equation}
  \item the gauge field $A_{\mu}$ is chosen as
\begin{equation}
	A_{\mu}\mathrm{d}x^{\mu}=A(r)\mathrm{d}t-P\cos{\theta}\mathrm{d}\varphi.
\end{equation}
\end{enumerate}
Here, $q$ is the linear coefficient of time and $P$ is a constant.

The field equations given by variation of the action can be solved with the above assumptions. Indeed, several solutions were found in ref. \cite{Babichev2015JCAP5.31}. Among them, one phenomenologically interesting case is $\gamma=0$, where the Maxwell field is not coupled to the scalar Galileon field. With additional assumption $\eta=\Lambda=0$, a perturbative solution can be found as \cite{Babichev2015JCAP5.31}
\begin{eqnarray}
\label{eq:mtrsol}
\begin{split}
h(r) & = 1 -\dfrac{\mu}{r}+\dfrac{\Gamma}{ r^2},\\
f(r) & =  h(r)\bigg(1+\dfrac{\Gamma}{r^2 }\bigg),
\end{split}
\end{eqnarray}
where $\Gamma$ is defined as
\begin{equation}
\Gamma=\dfrac{P^2+Q^2}{2(3\epsilon q^2-2)},
\end{equation}
and it is assumed that the magnetic charge $P$ and the electric charge $Q$ are small so that $|\Gamma|\ll r^2$. It can be checked that the metric \eqref{eq:mtrsol} is asymptotically flat and it returns to the Schwarzschild one  when $\Gamma=0$. For the case of $\Gamma>0$, the metric \eqref{eq:mtrsol} looks like the RN metric but differs from the RN metric because of $h(r)\neq f(r)$. \red{When $\Gamma<0$, the metric \eqref{eq:mtrsol} has some similarity with the tidal RN metric \cite{Dadhich2000PLB487.1} which is a black hole solution in the braneworld paradigm, but it also differs from the tidal RN one by $h(r)\neq f(r)$.} In the following sections, we will study the strong field lensing by the spacetime of the metric \eqref{eq:mtrsol} and investigate whether its observables can tell difference from other black holes.

%%%%%%%%%%%%%%%%%%%%%%%%%%%%%%%%%%%%
%%%%%SECTION3%%%%%%%%%%%%%%%%%%%%%%%%%
%%%%%%%%%%%%%%%%%%%%%%%%%%%%%%%%%%%%

\section{Lens equation}
\label{sec3}

In order to study the strong gravitational lensing by the charged Galileon black hole, we need a lens equation first to define the geometrical relationships among the observer, the lens and the source. The exact lens equation in an arbitrary Lorentzian spacetime can be found in refs. \cite{Frittelli1999PRD59.124001,Perlick2004LRR7.9}, and some works have given the explicit cases for spherically symmetric and static lensing \cite{Perlick2004PRD69.064017} and Schwarzchild lensing \cite{Frittelli2000PRD61.064021} without any requirement for flat background or hypothesis of the positions of the observer and the sources.

However, for the purpose of obtaining straightforward connections between models and observations as well as a more clear physical picture, we may put some approximations and hypotheses into the lens equation. The most reasonable and easy-doing approximation is the asymptotic approximation. It assumes the observer and the source are in a flat spacetime, while the curved spacetime only affects the deflection angle in the vicinity of the lens. Therefore, any other angular and distance quantities can be measured by the Euclidean geometry. It is worth mentioning that the asymptotic approximation also requires that (i) both the observer and the source should be far enough from the lens and (ii) the spacetime of the lens is asymptotically flat. In order to simplify the problem further, it can be assumed that the source lies behind the lens. Generally, even in the situations that the source lies between the observer and the lens or the source lies at the back of the observer, relativistic images can still exist, which is called retrolensing \cite{Eiroa2004PRD69.063004,Eiroa2005PRD71.083010}; and the assumption that the source is far from the lens can also be removed \cite{Bozza2007PRD76.083008}.

\begin{figure}[tbp]
\centering % \begin{center}/\end{center} takes some additional vertical space
\includegraphics[width=.9\textwidth]{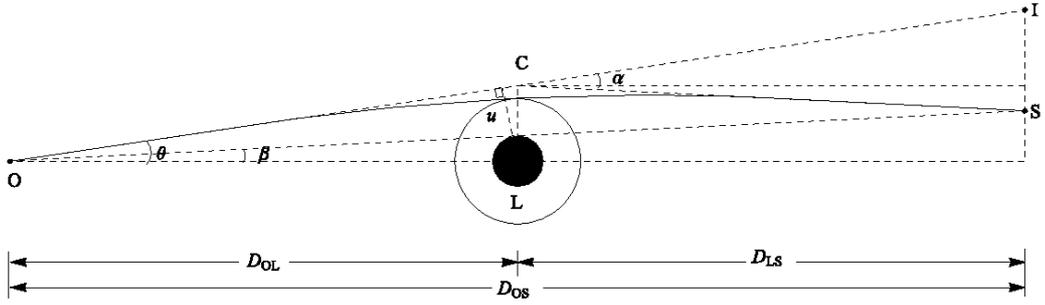}
% "\includegraphics" is very powerful; the graphicx package is already loaded
\caption{\label{Pic1} A schematic diagram for the geometry of the observer, the lens, the source and the image. O, L, and S respectively refer to the positions of the observer, the lens and the source. The light emitted from S is deflected by L then arrives at O. I is the position of an observed image and the two asymptotic line of the lights on the side of O and S meet at C. $\beta$ is the angular separation between the source and the lens; $\theta$ is the angular separation between the image and the lens; $\alpha$ is the deflection angle. $D_{\mathrm{OL}}$, $D_{\mathrm{LS}}$ and $D_{\mathrm{OS}}$ are the projected distance of observer-lens, lens-source and observer-source along the optical axis. The impact parameter $u$ is the distance of L to the line OI.}
\end{figure}

In this work, we focus on a conventional case that the source lies behind the lens. Under the asymptotic approximation and the hypothesis of the source position, \textcolor[rgb]{1,0,0}{we adopt the lens equation given by ref.~\cite{Bozza2001GRG33.1535} which is a practically feasible and widely used first order approximated form and reads as}
\begin{equation}
\label{eq:leqBZ}
\beta=\theta-\dfrac{D_{\mathrm{LS}}}{D_{\mathrm{OS}}}\Delta\alpha_n,
\end{equation}
where $\beta$ is the angular separation between the source and the lens; $\theta$ is the angular separation between the image and the lens; $\Delta\alpha_n=\alpha(\theta)-2n\pi$ is the extra angular deflection angle after a photon with a deflection angle $\alpha$ winding $n$ loops. $D_{\mathrm{LS}}$ and $D_{\mathrm{OS}}$ are the projected distance of lens-source and observer-source along the optical axis.  Their geometrical relationships are shown in figure \ref{Pic1}. The angular separation $\theta$ and the impact parameter $u$ can be easily exchanged by the relationship of $u=\tan{\theta}D_{\mathrm{OL}}\approx \theta D_{\mathrm{OL}}$.

The lens equation can also be defined in the way as ref. \cite{Ohanian1987AJP55.428}, which defines the position of deflection occurring by using the symmetry of the impact parameters of the observer side and the source side. Although it was assessed and found in ref. \cite{Bozza2008PRD78.103005} that this lens equation is a little more accurate for the strong field lensing than the asymptotically approximated lens equation \eqref{eq:leqBZ}, the latter one adopted in this work has the advantage in analysing the observational effects for its brief mathematical form.

%%%%%%%%%%%%%%%%%%%%%%%%%%%%%%%%%%%%
%%%%%SECTION4%%%%%%%%%%%%%%%%%%%%%%%%%
%%%%%%%%%%%%%%%%%%%%%%%%%%%%%%%%%%%%
\section{Gravitational lensing in the SDL}
\label{sec4}
Once we adopt the asymptotically approximated lens equation, the spacetime of the lens only affects the deflection angle $\alpha(\theta)$, which will be calculated in the SDL.

For convenience, we rewrite the metric \eqref{eq:mtrsol} of the charged Galileon black hole as
\begin{equation}
\label{eq:metric}
\mathrm{d}s^2=A(x)\mathrm{d}t^2-B(x)\mathrm{d}x^2-C(x)(\mathrm{d}\theta^2+\sin^2{\theta}\mathrm{d}\phi^2),
\end{equation}
where we take $\mu$ as the measure of distances and the functions are
\begin{eqnarray}
\label{eq:metricA}
A(x)&=&1-\frac{1}{x}+\frac{\Gamma}{x^2},\\
\label{eq:metricB}
B(x)&=&\left(1+\frac{\Gamma }{x^2}\right)^{-1}\cdot\left(1-\frac{1}{x}+\dfrac{\Gamma }{ x^2}\right)^{-1},\\
\label{eq:metricC}
C(x)&=&x^2.
\end{eqnarray}
The deflection angle for the null geodesic of a photon in the spacetime can be found as \cite{Weinberg1972Book,Virbhadra1998AA337.1}
\begin{equation}
\label{eq:dfagl}
\alpha(x_0)=-\pi+\int^{\infty}_{x_0}{2\sqrt{B(x)}\over{\sqrt{C(x)}\sqrt{{C(x)\over{C_0}}{A_0\over{A(x)}}-1}}}\mathrm{d}x,
\end{equation}
where $x_0$ is the closest distance of the photon to the black hole, $A_0$ and $C_0$ are the corresponding values of $A(x)$ and $C(x)$ at $x=x_0$. Substituting eqs. \eqref{eq:metricA}, \eqref{eq:metricB} and \eqref{eq:metricC} into eq.~\eqref{eq:dfagl}, we can obtain the exact deflection angle of the charged Galileon black hole.

However, the integral in eq.~\eqref{eq:dfagl} cannot be worked out in an explicit form. In the weak gravitational field, the deflection angle is a small angle and so an approximate solution can be obtained in the weak deflection limit (WDL). When dealing with the lensing in the strong gravitational field, the classical WDL is invalid. For solving this problem in the strong field, there are two feasible ways. One is seeking proper special functions to replace the integral \cite{Darwin1959PRSLSA249.180,Eiroa2005PRD71.083010}. The other way is expanding the integral in the SDL near the photon sphere \cite{Virbhadra2000PRD62.084003,Claudel2001JMP42.818}. The SDL method is valuable not only in providing an analytic representation of the deflection angle, but also in physically showing behavior of photons near the photon sphere. The WDL and SDL formulae work well in their own territories. Some strategies have been proposed to unify the expressions of the deflection angel from the photon sphere to infinity \cite{Amore2006PRD73.083004,Amore2006PRD74.083004,Amore2007PRD75.083005,Iyer2007GRG39.1563}. In the present work, we adopt the SDL method since the strong field lensing is focused on only.

In order to find the deflection angle in the SDL, we define the radius of the photon sphere $x_m$ as \cite{Virbhadra2000PRD62.084003,Claudel2001JMP42.818}
\begin{equation}
\label{eq:phosph}
\dfrac{C'(x)}{C(x)}=\dfrac{A'(x)}{A(x)},
\end{equation}
which can be solved as
\begin{equation}
x_{m}=\dfrac{1}{4} \left(3+\sqrt{9-32 \Gamma }\right).
\end{equation}
Following the approach proposed in ref. \cite{Bozza2002PRD66.103001}, we can have the deflection angle in the SDL as
\begin{equation}
\label{eq:alpha}
\alpha(\theta)=-\bar{a}\log\left(\dfrac{u}{u_m}-1\right)+\bar{b}+\mathcal{O}(u-u_m),
\end{equation}
where the coefficients $\bar{a}$ and $\bar{b}$ are
\begin{eqnarray}
\bar{a}&=&\frac{2 \sqrt{2} x_m}{\sqrt{8 x_m^2+6 x_m-9}},\\[1.5em]
\bar{b}&=&-\pi +\log (6)+\log (36)-4 \tanh ^{-1}\left(\frac{1}{\sqrt{3}}\right)\nonumber\\
&&+\left\{\dfrac{2}{9} \left[4+\log (6)\right]+\frac{4}{27} \left[4 \sqrt{3}-15+3 \log \left(-6 \sqrt{3}+12\right)\right]\right\}{\Gamma}+\mathcal{O}(\Gamma^2)\nonumber\\
&=&-0.4002 + 0.3023 {\Gamma}+\mathcal{O}(\Gamma^2),
\end{eqnarray}
and the impact parameter $u$ is given by \cite{Weinberg1972Book,Virbhadra2000PRD62.084003}
\begin{equation}
u=\sqrt{\dfrac{C_0}{A_0}},
\end{equation}
and its value at the photon sphere $x_0=x_m$ is
\begin{equation}
  \label{}
u_m = \frac{2 x_m^{3/2}}{\sqrt{2 x_m-1}}.
\end{equation}
The details of these calculations can be found in appendix~\ref{SecA.1}.

Besides the deflection angle, time delays between relativistic images are potential observables. The time taken by a photon from the source to the observer can be decomposed into three parts \cite{Bozza2004GRG36.435}
\begin{equation}
  \label{eq:T3parts}
  T =  \tilde{T}(x_0) - \int^{\infty}_{D_{\mathrm{OL}}}\bigg|\frac{\mathrm{d}t}{\mathrm{d}x}\bigg|\mathrm{d}x - \int^{\infty}_{D_{\mathrm{LS}}}\bigg|\frac{\mathrm{d}t}{\mathrm{d}x}\bigg|\mathrm{d}x,
\end{equation}
where $\tilde{T}(x_0)$ is defined as \cite{Weinberg1972Book,Bozza2004GRG36.435}
\begin{equation}
\label{eq:dftime}
\tilde{T}(x_0)=\int^{\infty}_{x_0}{2\sqrt{B(x)C(x)A_0}\over{A(x)\sqrt{C_0}\sqrt{{C(x)\over{C_0}}{A_0\over{A(x)}}-1}}}\mathrm{d}x,
\end{equation}
The last two terms in eq. \eqref{eq:T3parts} can be easily handled since the photon is far away from the black hole. But it is not the case for the first term in eq. \eqref{eq:T3parts} because its integral is divergent at $x_0$. We can also apply the procedure proposed in ref. \cite{Bozza2004GRG36.435} to deal with it. The resulting formula has the same form as eq.~\eqref{eq:alpha} and it reads as \cite{Bozza2004GRG36.435}
\begin{equation}
\label{eq:time}
\tilde{T}(u)=-\tilde{a}\ln{\left(\dfrac{u}{u_m}-1\right)}+\tilde{b}+\mathcal{O}(u-u_m),
\end{equation}
where $\tilde{a}$ and $\tilde{b}$ are the coefficients for the SDL. It was found in ref. \cite{Bozza2004GRG36.435} that for a spherically symmetric metric, there is a very important relation as
\begin{equation}
  \label{}
  \frac{\tilde{a}}{\bar{a}} = u_m.
\end{equation}
See appendix~\ref{SecA.2} for more details.

%%%%%%%%%%%%%%%%%%%%%%%%%%%%%%%%%%%%%%%%%%%
%%%%%%%%SECTION5%%%%%%%%%%%%%%%%%%%%%%
%%%%%%%%%%%%%%%%%%%%%%%%%%%%%%%%%%%%%%%%
\section{Observables}

\label{sec5}

After the lens equation \eqref{eq:leqBZ} as well as the deflection angle \eqref{eq:alpha} and the time delay \eqref{eq:time} in the SDL are obtained, we can calculate the observables of the strong field lensing by the charged Galileon black hole, including the positional separations, brightness differences and time delays between relativistic images.

%%%%%%%%%%%%%%%%%%%%%%%%%%%%%%%%%%%%%%%%
%%%%%%%%%%%%%%%%%%%%%%%%%%%%%%%%%%
\subsection{Angular separation and brightness difference of relativistic images}
\label{sec42}

By using the lens equation \eqref{eq:leqBZ}, we have established a connection between the true position of the source $\beta$ and the apparent position of the image $\theta$. The deflection angle $\alpha(\theta)$ in that equation is expressed by the SDL coefficients: $u_m$, $\bar{a}$ and $\bar{b}$ in eq.~\eqref{eq:alpha}, which can be calculated according to the metric of the black hole (see appendix~\ref{SecA.1} for details). Hence the position of a $n$-loop relativistic image $\theta_n$ can be expressed by a function of  $u_m$, $\bar{a}$ and $\bar{b}$. Taking eq.~\eqref{eq:alpha} into eq.~\eqref{eq:leqBZ}, we can have \cite{Virbhadra2000PRD62.084003,Bozza2002PRD66.103001}
\begin{equation}
  \label{}
  \label{eq:theta0}
  \theta_n = \theta_n^0+\Delta\theta_n,
\end{equation}
where
\begin{eqnarray}
\label{eq:theta1}
\theta^0_n&=&\dfrac{u_m}{D_{\mathrm{OL}}}\bigg\{1+\exp\bigg[\frac{\bar{b}-2n\pi}{\bar{a}}\bigg]\bigg\},\\
\Delta\theta_n&=&\dfrac{u_m(\beta-\theta_n^0)D_{\mathrm{OS}}}{\bar{a}D_{\mathrm{LS}}D_{\mathrm{OL}}}\exp\bigg[\frac{\bar{b}-2n\pi}{\bar{a}}\bigg].
\end{eqnarray}
Here $\theta^0_n$ is the corresponding value of $\theta$ when $\alpha(\theta^0_n)=2n\pi$. The corrected term $\Delta\theta_n$ are much smaller than the main term $\theta_n^0$.

Besides the positions, another important observable is the brightness or the magnification of the images. The magnification of the $n$-loop relativistic image's brightness from the original source's brightness is given by \cite{Refsdal1964MNRAS128.295,Liebes1964PR133.835}
\begin{equation}
\mu_n=\dfrac{1}{(\beta/\theta)\partial\beta/\partial\theta}\bigg|_{\theta_n^0}.
\end{equation}
If we suppose that the outermost image could at least be able to separated from the inner packed others and assume $\beta\sim\theta_\infty$ and $\bar{a}\sim1$, we can have three observables as \cite{Bozza2002PRD66.103001}
\begin{eqnarray}
\theta _{\infty }&=&\frac{u_m}{D_{\text{OL}}},\\
s&=&\theta_1-\theta_\infty=\theta _{\infty } \exp \left(\frac{\bar{b}}{\bar{a}}-\frac{2 \pi }{\bar{a}}\right),\\[0.5em]
r&=&2.5 \log _{10}\left(\dfrac{\mu_1}{\sum^\infty_{n=2}\mu_n}\right)=2.5 \log _{10}\left[\ \exp \left(\frac{2 \pi }{\bar{a}}\right)\ \right],
\end{eqnarray}
where $\theta_\infty$ is the asymptotic position approached by a set of images in the limit $n\rightarrow\infty$, $s$ is the angular separation between  the outermost image ($n=1$) and the packed others ($n=2,3,\cdots\infty$), and $r$ is the magnitude difference between the outermost image and the packed images. If the observables are available, the coefficients in the SDL can be obtained by \cite{Bozza2002PRD66.103001}
\begin{eqnarray}
u_m&=&\theta _{\infty }D_{\text{OL}},\\[0.5em]
\bar{a}&=&\dfrac{5\pi}{r\log{10}}\\
\bar{b}&=&\bar{a}\left[\ \log{ \left(\dfrac{s}{\theta_{\infty}}\right)}+r\dfrac{2}{5}\log{10}\ \right].
\end{eqnarray}
Thus, we have a bidirectional map between the observables $\theta _{\infty }$, $s$, $r$ and the model coefficients $u_m$, $\bar{a}$, $\bar{b}$.

\begin{figure}[tbp]
\centering
\includegraphics[width=1\textwidth]{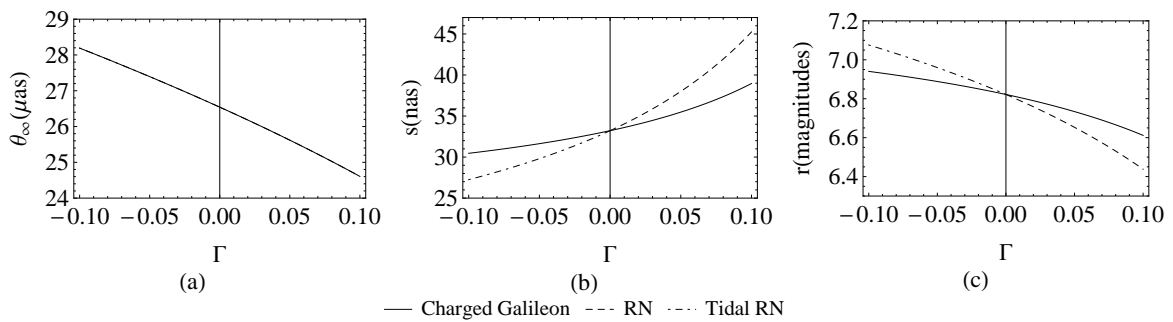}
\vspace{-0.6cm}
\caption{\label{Pic2} The estimated observables $\theta_\infty$, $s$ and $r$ of the charged Galileon black hole, the RN black hole \red{and the tidal RN black hole}. It is assumed that they have the same mass and distance as Sgr A*. Diagram (a), (b) and (c) represent $\theta_\infty$, $s$ and $r$ against the parameter $\Gamma$, where the solid lines refer to the charged Galileon black hole, \red{the dashed and dot dashed lines refer to the RN black hole and the tidal RN black hole when the tidal charge parameter $\tilde{Q}=\Gamma$. The charged Galileon, the RN and the tidal black holes have identical $\theta_{\infty}$ so that their curves coincide with each other in diagram (a).}}
\end{figure}

\begin{figure}[htbp]
\centering
\includegraphics[width=1\textwidth]{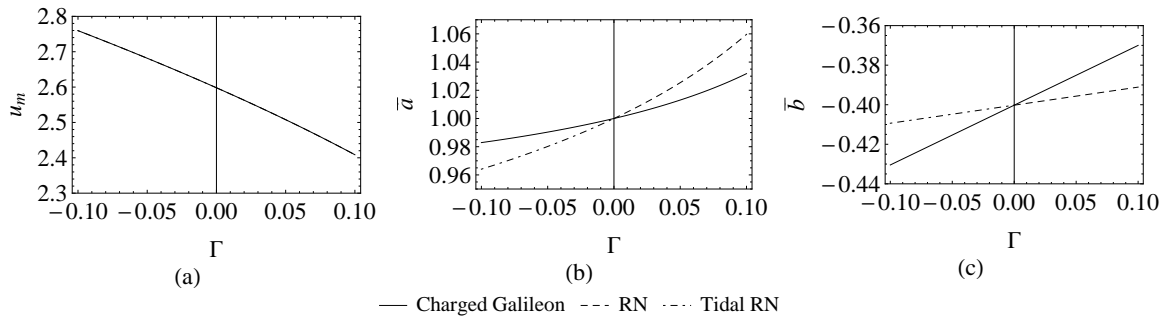}
\vspace{-0.6cm}
\caption{\label{Pic3} The SDL coefficients $u_m$, $\bar{a}$, $\bar{b}$ of the charged Galileon black hole, the RN black hole \red{and the tidal RN black hole}. These black holes are also assumed to have the same mass and distance as Sgr A*. Diagram (a), (b) and (c) represent the SDL coefficients against the parameter $\Gamma$, where the solid lines refer to the charged Galileon black hole, \red{the dashed and dot dashed lines refer to the RN black hole and the tidal RN black hole where the tidal charge parameter is set as $\tilde{Q}=\Gamma$. The charged Galileon, the RN and the tidal black holes have identical $u_m$ so that their curves coincide with each other in diagram (a).}}
\end{figure}

Table \ref{Tab1} shows the estimated observables $\theta _{\infty}$, $s$, $r$ and the SDL coefficients $u_m$, $\bar{a}$, $\bar{b}$ for the charged Galileon black hole like the supermassive black hole Sgr A* in our galaxy with $M_{\bullet}=4.31\times10^6$ $M_\odot$ and \mbox{$D_{\mathrm{OL}}= D_{\bullet}=8.33$ kpc} \cite{Gillesse2009ApJ707.L114}. We also estimate these quantities for the Schwarzschild black hole, the RN black hole, \red{as well as the tidal RN black hole} for comparison. \red{The tidal RN metric is a special solution of black holes in the braneworld paradigm given by ref. \cite{Dadhich2000PLB487.1} (see ref. \cite{Maartens2004LRR7.7} for a review on the braneworld gravity). Several works have been done to study the strong lensing effects of the tidal RN black hole \cite{Bin-Nun2010PRD81.123011,Bin-Nun2010PRD82.064009,Horvath2013AN334.1047,Whisker2005PRD71.064004,Zakharov2014PRD90.062007}. The RN metric and the tidal RN metric have the same formality and the $r^{-2}$ terms in both metrics are controlled by the tidal charge parameter $\tilde{Q}$. $\tilde{Q}$ in the RN metric is positive and equal to the square of the electric charge of the black hole, while $\tilde{Q}$ in the tidal RN metric can be negative due to the gravitational effects from the fifth dimension \cite{Dadhich2000PLB487.1}}. It is found  in our estimation that, for the charged Galileon black hole, $\theta _{\infty }$ is at the level of $\sim 26$ $\mu$as while $s$ is much smaller at the level of $\sim 34$ nano-arcsecond (nas). Figures \ref{Pic2} and \ref{Pic3} show how the observables and the coefficients change against different lenses.

\begin{table}
\centering
\scriptsize
\tabcolsep 1.95mm
\begin{tabular*}{1\textwidth}{lccccccccc}
\hline
\hline
&Schwarzschild&\multicolumn{4}{c}{Charged Galileon}&\multicolumn{2}{c}{Tidal RN}&\multicolumn{2}{c}{RN}\\
&&$\Gamma\!=\!-0.1$&$\Gamma\!=\!-0.05$&$\Gamma\!=\!0.05$&$\Gamma\!=\!0.1$&$\tilde{Q}\!=\!-0.1$&$\tilde{Q}\!=\!-0.05$&$\tilde{Q}\!=\!0.05$&$\tilde{Q}\!=\!0.1$\\
\hline
$\theta_{\infty}$ ($\mu$as)&26.54	&	28.19	&	27.39	&	25.62	&	24.61	&	28.19	&	27.39	&	25.62	&	24.61	\\
$s$ (nas)&33.22	&	30.46	&	31.63	&	35.48	&	38.96	&	27.25	&	29.81	&	37.98	&	45.25	\\
$r$ (mag)&6.822	&	6.940	&	6.888	&	6.734	&	6.612	&	7.076	&	6.960	&	6.653	&	6.438	\\
$u_m$ ($R_s$)&2.598	&	2.760	&	2.682	&	2.508	&	2.409	&	2.760	&	2.682	&	2.508	&	2.409	\\
$\bar a$&1	&	0.9829	&	0.9903	&	1.013	&	1.032	&	0.9641	&	0.9802	&	1.025	&	1.060	\\
$\bar b$&-0.4002	&	-0.4305	&	-0.4153	&	-0.3851	&	-0.3700	&	-0.4094	&	-0.4048	&	-0.3957	&	-0.3911	\\
\hline
\end{tabular*}
\caption{\label{Tab1}The estimated observables $\theta _{\infty }$, $s$, $r$ and the SDL coefficients $u_m$, $\bar{a}$, $\bar{b}$ for the Schwarzschild, the charged Galileon, the tidal RN and the RN black holes. We assume these black holes have the same mass and distance as Sgr A* with $M_{\bullet}=4.31\times10^6$ $M_\odot$ and $D_{\mathrm{OL}}= D_{\bullet}=8.33$ kpc \cite{Gillesse2009ApJ707.L114}. \textcolor[rgb]{1,0,0}{$\tilde{Q}$ is the tidal charge parameter which has a positive value for the RN black hole and has a negative value for the tidal RN black hole.} $\theta_{\infty}$ and $s$ are respectively in the units of micro-arcsecond ($\mu$as) and nano-arcsecond (nas). The unit of $u_m$ is Schwarzschild radius $R_s=2GM_{\bullet}/c^2$. $r$, $\bar a$ and $\bar b$ are dimensionless.}
\end{table}

The metric coefficients $A(x)$ and $C(x)$ of a charged Galileon black hole, a RN black hole \red{and a tidal RN black hole} have the same structure, i.e., $\Gamma$ can effectively be equivalent to $\tilde{Q}$, and difference only appears in $B(x)$. It is of interest whether observation of strong field lensing can distinguish one from the other. As table \ref{Tab1} shown, when $\Gamma>0$, $\theta_{\infty}$ of the charged Galileon black hole is always smaller than the one of the Schwarzschild black hole with the same masses, the difference can reach $1.93$ $\mu$as when $\Gamma=0.1$. The charged Galileon and the RN black holes have identical $\theta_{\infty}$ but slightly different $s$ and $r$. It means, if the outermost relativistic image is unable to be separated and all of the images are packed together, then the strong field lensings of these two kinds of black holes look the same. But once we are able to achieve the demanded resolution, we will find the first image is closer to the packed one and the brightness difference between the two images is larger in the vicinity of the charged Galileon black hole than those nearby the RN black hole; see figure \ref{Pic2} (b) and (c). When $\Gamma=\tilde{Q}=0.1$ for the charged Galileon and RN black holes, the differences of their $s$ and $r$ are respectively $\sim6$ nas and $\sim0.2$ magnitude.

In the case of $\Gamma<0$, $\theta_{\infty}$ of the charged Galileon black hole is always larger than the one of the Schwarzschild black hole with the same masses. When $\Gamma=-0.1$, the difference of $\theta_{\infty}$ between the two is $1.65$ $\mu$as. \red{When $\Gamma$ and $\tilde{Q}$ are negative, $\theta_{\infty}$ is same for the charged Galileon black hole and the tidal RN black hole, but the charge Galileon black hole has larger $s$ and smaller $r$. When $\Gamma=\tilde{Q}=-0.1$, the values of their differences are respectively $\sim3$ nas and $\sim0.1$ magnitude. In addition, if the first relativistic image can be resolved, we will find that the charged Galileon black hole can generate a larger separation and weaker brightness difference between the first relativistic image and the packed others than the tidal RN black hole does; see figure \ref{Pic2} (b) and (c)}.

Separating the first relativistic image from the packed others is a grand challenge for observations, since the angular separation between them $s$ is extremely small although the brightness difference is theoretically notable. For the closet supermassive black hole Sgr A*, $s$ is only $\sim 30$ nas, which is far beyond the limit of current technology. \red{The differences of $s$ among the Galileon, RN and tidal RN black holes range from $\sim3$ to $\sim7$ nas, which makes the task of discriminating the charged Galileon black hole via observation much more difficult.}

\subsection{Time delays between relativistic images}

Another important kind of observables are time delays among the relativistic images. If the observer, the lens and the source are aligned, the time delay between a $n$-loop and a $m$-loop relativistic image is \cite{Bozza2004GRG36.435}
\begin{equation}
  \label{}
  \label{eq:TD}
\Delta T_{n,m} = \Delta T_{n,m}^{\ 0}+\Delta T_{n,m}^{\ 1},
\end{equation}
where
\begin{eqnarray}
\label{eq:TD0}
\Delta T_{n,m}^{\ 0}&=&2 \pi (n-m) u_m,\\[0.5em]
\label{eq:TD1}
\Delta T_{n,m}^{\ 1}&=&2\sqrt{\frac{B_m}{A_m}} \sqrt{\dfrac{u_m}{\hat c}}\exp\left({\frac{\bar{b}}{2\bar{a}}}\right)\left[\exp\left({-\frac{m\pi}{\bar{a}}}\right)-\exp\left({-\frac{n\pi}{\bar{a}}}\right)\right].
\end{eqnarray}
Here, $A_m$ and $B_m$ are the values of $A(x)$ and $B(x)$ at $x=x_m$, $\hat c$ is the coefficient in the approximated formula $u-u_m=\hat c(x_0-x_m)^2$. $\Delta T_{n,m}^{\ 0}$ is the main term of time delay, while $\Delta T_{n,m}^{\ 1}$ is its much smaller correction, i.e., $\Delta T_{n,m}^{\ 1}\ll\Delta T_{n,m}^{\ 0}$. More details can be found in appendix~\ref{SecA.2}.

Table \ref{Tab.2} and figure \ref{Pic4} show the estimated time delays between the relativistic images of the charged Galileon black hole, the Schwarzchild black hole, the RN black hole \red{and the tidal RN black hole}, which are assumed to have the mass $M_{\bullet}$ and the distance $D_{\bullet}$.  $\Delta T_{n,m}^{\ 0}$ and $\Delta T_{n,m}^{\ 1}$ are both represented in the unit of $2GM_{\bullet}/c^{3}\approx42.45$ s. Based on the values of $m$ and $n$, we consider six different cases: ($n=2$, $m=1$), ($n=3$, $m=2$), ($n=4$, $m=3$), ($n=3$, $m=1$),($n=4$, $m=2$) and ($n=4$, $m=1$). It is clearly shown that the time delay grows with the increment of the difference between $n$ and $m$.

As discussed previously, there is a potentially important question whether the observations of time delay can distinguish the charged Galileon black hole from the RN black hole for $\Gamma>0$ \red{and from the tidal RN black hole for $\Gamma<0$}. According to eq. \eqref{eq:TD}, the time delay consists of two contributions. The main term $\Delta T_{n,m}^{\ 0}$ only depends on $u_m$, whose values are identical for the charged Galileon, the RN and the \red{tidal RN} black holes if $\tilde{Q}=\Gamma$ (see table \ref{Tab1}). The only difference comes from the second term $\Delta T_{n,m}^{\ 1}$, but it is smaller than the main term by 2 to 4 orders of magnitude. Figure \ref{Pic4} (b) shows the second term of time delay between different relativistic images and figure \ref{Pic4} (c) shows the ratio of the second term to the total time delay. We find that $\Delta T_{n,m}^{\ 1}$ is only notable in the cases of $m=1$, among which the most significant one is $\Delta T_{2,1}^{\ 1}$. It means that if the first relativistic image and the second one can be separated, the time delay between them might provide a chance to distinguish a charged Galileon black hole from a RN black hole or \red{a tidal RN black hole}. As table \ref{Tab.2} shown, in order to discriminate the charged Galileon black hole from other black holes by the most significant term $\Delta T_{2,1}$, the detection of $\Delta T_{2,1}$ needs to have an accuracy better than the level of $\sim2\times10^{-2}$, which corresponds to the level of about 1 s for Sgr A*. In fact, it is still difficult to observe the telltale term in the time delay even if the differences are evident because the first and second relativistic images need to be separated firstly.

\begin{table}
\centering
\scriptsize
\tabcolsep 1.95mm
\begin{tabular}{lccccccccc}
\hline
\hline
Time &Schwarz-&\multicolumn{4}{c}{Charged Galileon}&\multicolumn{2}{c}{Tidal RN}&\multicolumn{2}{c}{RN}\\
delay&schild&$\Gamma\!=\!-0.1$&$\Gamma\!=\!-0.05$&$\Gamma\!=\!0.05$&$\Gamma\!=\!0.1$&$\tilde{Q}\!=\!-0.1$&$\tilde{Q}\!=\!-0.05$&$\tilde{Q}\!=\!0.05$&$\tilde{Q}\!=\!0.1$\\
\hline\\[-0.5em]
$\Delta T_{2,1}$&16.57	&	17.58	&	17.09	&	16.01	&	15.40	&	17.56	&	17.08	&	16.02	&	15.43	\\
$\Delta T_{3,2}$&16.33	&	17.35	&	16.86	&	15.77	&	15.15	&	17.35	&	16.86	&	15.77	&	15.15	\\
$\Delta T_{4,3}$&16.32	&	17.34	&	16.85	&	15.76	&	15.13	&	17.34	&	16.85	&	15.76	&	15.13	\\
$\Delta T_{3,1}$&32.91	&	34.93	&	33.95	&	31.78	&	30.55	&	34.91	&	33.94	&	31.79	&	30.58	\\
$\Delta T_{4,2}$ &32.66	&	34.69	&	33.71	&	31.53	&	30.28	&	34.69	&	33.71	&	31.53	&	30.28	\\
$\Delta T_{4,1}$&49.23	&	52.27	&	50.80	&	47.54	&	45.68	&	52.25	&	50.79	&	47.55	&	45.71	\\
\\[-0.5em]
\hline
\\[-0.5em]
$\Delta T_{2,1}^{\ 1}$&0.2487	&	0.2419	&	0.2445	&	0.2554	&	0.2664	&	0.2249	&	0.2353	&	0.2670	&	0.2936	\\
$\Delta T_{3,2}^{\ 1}$ ($10^{\!-\!1}$) &0.1075	&	0.09896	&	0.1025	&	0.1149	&	0.1268	&	0.08647	&	0.09541	&	0.1247	&	0.1514	\\
%\tiny{ ($\times10^{\!-\!1}$)}&&&&&&&&&\\
$\Delta T_{4,3}^{\ 1}$ ($10^{\!-\!3}$)  &0.4645	&	0.4049	&	0.4295	&	0.5172	&	0.6038	&	0.3324	&	0.3869	&	0.5826	&	0.7809	\\
%\tiny{($\times10^{\!-\!3}$)}&&&&&&&&&\\
$\Delta T_{3,1}^{\ 1}$&0.2595	&	0.2518	&	0.2548	&	0.2669	&	0.2791	&	0.2336	&	0.2448	&	0.2795	&	0.3087	\\
$\Delta T_{4,2}^{\ 1}$ ($10^{\!-\!1}$)  &0.1121	&	0.1030	&	0.1068	&	0.1201	&	0.1329	&	0.08979	&	0.09928	&	0.1306	&	0.1592	\\
%\tiny{($\times10^{\!-\!1}$)}&&&&&&&&&\\
$\Delta T_{4,1}^{\ 1}$&0.2599	&	0.2522	&	0.2552	&	0.2674	&	0.2797	&	0.2339	&	0.2452	&	0.2801	&	0.3095	\\
\\[-0.5em]
\hline
\\[-0.5em]
$\tilde\eta_{2,1}$&-1.8	&	-1.9	&	-1.8	&	-1.8	&	-1.8	&	-1.9	&	-1.9	&	-1.8	&	-1.7	\\
$\tilde\eta_{3,2}$&-3.2	&	-3.2	&	-3.2	&	-3.1	&	-3.1	&	-3.3	&	-3.2	&	-3.1	&	-3.0	\\
$\tilde\eta_{4,3}$&-4.5	&	-4.6	&	-4.6	&	-4.5	&	-4.4	&	-4.7	&	-4.6	&	-4.4	&	-4.3	\\
$\tilde\eta_{3,1}$&-2.1	&	-2.1	&	-2.1	&	-2.1	&	-2.0	&	-2.2	&	-2.1	&	-2.1	&	-2.0	\\
$\tilde\eta_{4,2}$&-3.5	&	-3.5	&	-3.5	&	-3.4	&	-3.4	&	-3.6	&	-3.5	&	-3.4	&	-3.3	\\
$\tilde\eta_{4,1}$&-2.3	&	-2.3	&	-2.3	&	-2.2	&	-2.2	&	-2.3	&	-2.3	&	-2.2	&	-2.2	\\
\\[-0.5em]
\hline
\end{tabular}
\caption{\label{Tab.2}The estimated time delays between the outermost four relativistic images for black holes with the mass $M_{\bullet}$ and the distance $D_{\bullet}$. $\Delta T_{n,m}$ is the total time delay between the $m$-loop image and the $n$-loop image and $\Delta T_{n,m}^{\ 1}$ is the correction on its main term. $\Delta T_{n,m}$ and $\Delta T_{n,m}^{\ 1}$ are both represented in the unit of $2GM_{\bullet}/c^{3}\approx42.45$ s. $\tilde\eta_{n,m}=\log_{10}{(\Delta T_{n,m}^{\ 1}/\Delta T_{n,m})}$ is the logarithmic ratio of them.}
\end{table}

\begin{figure}[htbp]
\centering
\includegraphics[width=1.03\textwidth,trim=10 5 10 0,clip]{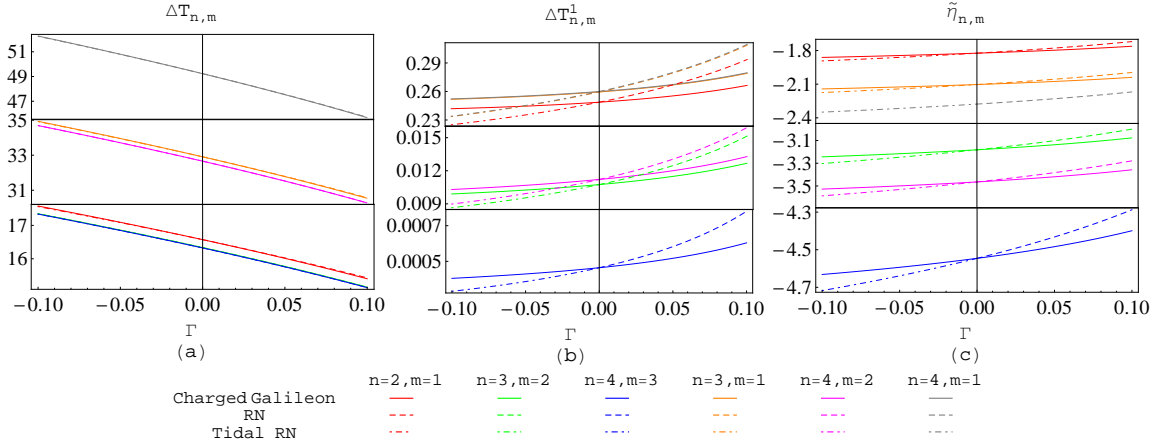}
\caption{\label{Pic4}  Diagrams (a), (b) and (c) show $\Delta T_{n,m}$, $\Delta T_{n,m}^{\ 1}$ and $\tilde\eta_{n,m}$ of the charged Galileon black hole, the RN black hole and \red{the tidal RN black hole}  with respect to $\Gamma$. These black holes are also assumed to have the mass $M_{\bullet}$ and the distance $D_{\bullet}$. The solid lines refer to the charged Galileon black hole, the dashed and dot dashed lines represent the RN black hole \red{and the tidal RN black hole} when the tidal charge parameter $\tilde{Q}=\Gamma$. The color of the lines and the corresponding values of $m$ and $n$ can be seen in the legend chart.}
\end{figure}

%%%%%%%%%%%%%%%%%%%%%%%%%%%%%%%%%%%%%%%%%%%
%%%%%%%%%SECTION6%%%%%%%%%%%%%%%%%%%%%
%%%%%%%%%%%%%%%%%%%%%%%%%%%%%%%%%%%%%%%%%%%
\section{Conclusions and discussion}
\label{sec6}

In this work we analyse the strong field gravitational lensing effects caused by the charged Galileon black hole. It is possible for the charged Galileon black hole to evade the no-hair theory. The strong field lensing can provide an opportunity to test it in the vicinity of the black hole by observing a set of infinite discrete relativistic images near the photon sphere. Those observations might be able to achieve in the future.

We adopt the asymptotic lens equation to describe the geometrical relationships among the observer, the lens, the source and the images, and suppose the source lies behind the lens. Then we expand the deflection angle on the neighborhood of photon sphere and get the SDL coefficients, so all the observables, including the angular separations, brightness differences and time delays between the relativistic images, can be expressed by these coefficients. We estimate these observables by taking Sgr A* as an example and make comparisons among the charged Galileon black hole, schwarzschild black hole, the RN black hole \red{and the tidal RN black hole}. We find that when $\Gamma>0$, it is difficult to distinguish the charged Galileon black hole from the RN black hole. \red{When $\Gamma<0$, the observables generated by the charged Galileon black hole are close to those given by the tidal RN black hole}. It might be helpful if we can separate the outermost relativistic images and determine their differences on the positions, brightness difference and time delay, which requires the angular resolution better than the level of $30$ nas for Sgr A*. In order to distinguish the charged Galileon black hole from the RN black hole or \red{the tidal RN black hole}, the accuracy of the measured separation between the first image and the packed others needs to be better than about 3 nas; the photometric uncertainty has to better than $\sim 0.1$ mag; and the time delay detection is required to achieve the accuracy better than about 1 s. 

On the  possibility of observing the strong filed lensing, it is a challenge to achieve such a high resolution at present but the perspective of future observation is promising. The Event Horizon Telescope (EHT) \footnote{\url{http://www.eventhorizontelescope.org/}}, a global network of millimetre-wave very long baseline interferometric array, is expected to provide high-angular-resolution observation of Sgr A* and M87. It is claimed that the EHT will have the ability to discern things at the event horizon scale, access $\gtrsim10$ $\mu$as angular scale \cite{Broderick2014ApJ784.7}. However due to the limited number of baselines, the current EHT array might not be able to directly image the black hole, but provide the data in the Fourier domain, which can be fitted with given geometric model and accretion flow simulations \cite{Kamruddin2013MNRAS434.765,Ricarte2015MNRAS446.1973}. The scale and the shape of the shadow of the Sgr A* can be calculated more stringent in the future and may be able to testing gravitational theories \cite{Psaltis2008LRR11.9,Lacroix2013AA554.A36,Broderick2014ApJ784.7,Psaltis2015ApJ814.115}. But for relativistic images, there is still a long journey before detection. Therefore, it is necessary to give an intuitive view of what the observed images will look like through the analytical approach in the framework of GR and alternative theories of gravity.

\appendix

\section{Appendix}

\subsection{Calculating the deflection angle in the SDL}

\label{SecA.1}

In this appendix, we will present the details of the calculation for the deflection angle in the SDL by making use of the approach proposed in ref. \cite{Bozza2002PRD66.103001}. For a static and spherically symmetric black hole, its generic line element can be written as
\begin{equation}
	\label{appeq:ds2}
	\mathrm{d}s^2 = A(x)\mathrm{d}t^2-B(x)\mathrm{d}x^2-C(x)(\mathrm{d}\theta^2+\sin^2{\theta}\mathrm{d}\varphi^2),
\end{equation}
where, in order to correctly match asymptotic requirement of the lens equation, we assume the metric returns to the Schwarzchild solution as $x\to\infty$, i.e.,
\begin{eqnarray}
	\lim_{x\to\infty} A(x)&=&1-{{1}\over{x}},\nonumber\\
\lim_{x\to\infty} B(x) & = &1+{{1}\over{x}},\\
\lim_{x\to\infty} C(x)&=&x^2.\nonumber
\end{eqnarray}
Here $x$ is in the unit of the Schwarschild radius. In the case of a charged Galileon black hole, they are \cite{Babichev2015JCAP5.31}
\begin{eqnarray}
A(x)&=&1-\frac{1}{x}+\dfrac{\Gamma }{ x^2},\\
B(x)&=&\left(1+\frac{\Gamma }{x^2}\right)^{-1}\cdot\left(1-\frac{1}{x}+\dfrac{\Gamma }{ x^2}\right)^{-1},\\
C(x)&=&x^2,
\end{eqnarray}
where
\begin{equation}
\Gamma=\dfrac{P^2+Q^2}{2(3\epsilon q^2-2)}, \qquad|\Gamma|\ll x^2.
\end{equation}

For a null geodesic, it can be found as \cite{Bozza2004GRG36.435}
\begin{equation}
	{\mathrm{d}\varphi\over{\mathrm{d}x}}=P_1(x,x_0)P_2(x,x_0),\\
\end{equation}
where $x_0$ is the closest distance of the photon to the black hole and two functions are
\begin{eqnarray}
	P_1(x,x_0)&=&\frac{\sqrt{B(x)A(x)C_0}}{C(x)},\\
\label{eq:P2}
P_2(x,x_0)&=&{1\over{\sqrt{A_0-A(x){C_0\over{C(x)}}}}}.
\end{eqnarray}
The subscript $0$ of a quantity means its value at $x=x_0$.

The deflection angle is \cite{Weinberg1972Book,Virbhadra1998AA337.1}
\begin{equation}
	\alpha(x_0) =2\int^{\infty}_{x_0}{\mathrm{d}\varphi\over{dx}}\mathrm{d}x-\pi\equiv I(x_0)-\pi.
\end{equation}
We can also define some useful variables as \cite{Bozza2002PRD66.103001}
\begin{equation}
y=A(x),\quad y_0=A_0,
\end{equation}
and
\begin{equation}
z={y-y_0\over{1-y_0}}.
\end{equation}
The integral $I(x_0)$ can be reduced to \cite{Bozza2002PRD66.103001}
\begin{equation}
  \label{}
  \label{eq:A11}
I(x_0) = \int^1_0R(z,x_0)f(z,x_0)\mathrm{d}z,
\end{equation}
where
\begin{eqnarray}
R(z,x_0)&=&2\dfrac{(1-y_0)}{A'(x)}P_1(x,x_0),\\
f(z,x_0)&=&P_2(x,x_0).
\end{eqnarray}
Here, $R(z,x_0)$ is the regular term and $f(z,x_0)$ is the divergent term which diverges for $z\to0$.

The function $f(z,x_0)$ can be approximated as \cite{Bozza2002PRD66.103001}
\begin{equation}
f(z,x_0)\sim f_0(z,x_0)=\dfrac{1}{\sqrt{\hat\alpha z+\hat\beta z^2}},
\end{equation}
where
\begin{eqnarray}
\hat\alpha&=&{1-y_0\over{C_0 A_0'}} \left(y_0 C_0'-C_0 A_0'\right)\\
\hat\beta&=&{(1-y_0)^2\over{2 C_0^2 \left(A_0'\right)^3}}[2C_0C_0'A_0'^2+(C_0C_0''-2C_0'^2)y_0A_0'-C_0C_0'y_0A_0''].
\end{eqnarray}
For the charged Galileon black hole, the coefficients in $f_0(z,x_0)$ are
\begin{eqnarray}
\hat{\alpha}&=&\dfrac{\left(\Gamma -x_0\right) \left(4 \Gamma +2 x_0^2-3 x_0\right)}{x_0^2 \left(2 \Gamma -x_0\right)},\\
\hat{\beta}&=&-\dfrac{\left(\Gamma -x_0\right){}^2 \left[8 \Gamma ^2-9 \Gamma  x_0-\left(x_0-3\right) x_0^2\right]}{x_0^2 \left(2 \Gamma -x_0\right){}^3}.
\end{eqnarray}
We get the radius of the photon sphere $x_{m}$ by solving the equation $\hat{\alpha}=0$, and the result is
\begin{equation}
x_{m}=\dfrac{1}{4} \left(3+\sqrt{9-32 \Gamma }\right).
\end{equation}
Replacing $x_0$ with $x_{m}$ in $\hat{\beta}$, we have
\begin{equation}
\hat{\beta}_m=\dfrac{512 \Gamma ^3+\left(-80 \sqrt{9-32 \Gamma }-144\right) \Gamma ^2+\left(16 \sqrt{9-32 \Gamma }+32\right) \Gamma +3 \sqrt{9-32 \Gamma }-9}{64 (1-4 \Gamma )^2 \Gamma }.
\end{equation}
The subscript $m$ of a quantity means its value at $x=x_m$.

Then, the integral eq. \eqref{eq:A11} can be written by a summation of two integrals as \cite{Bozza2002PRD66.103001}
\begin{equation}
I(x_0)=I_D(x_0)+I_R(x_0),
\end{equation}
where $I_D$ is divergent and $I_R$ is regular, and they are
\begin{eqnarray}
	I_D(x_0)&=&\int^1_0 R(0,x_m)f_0(z,x_0)\mathrm{d}z,\\
I_R(x_0)&=&\int^1_0 \left[R(z,x_0)f(z,x_0)-R(0,x_m)f_0(z,x_0)\right]\mathrm{d}z.
\end{eqnarray}
For the charged Galileon black hole, it can be found that
\begin{equation}
R(0,x_m)=\frac{2 x_m \left(\Gamma -x_m\right) }{2 \Gamma -x_m}\sqrt{\frac{1}{\Gamma +x_m^2}}.
\end{equation}
These two integrals can be expand at the photo sphere neighbourhood as \cite{Bozza2002PRD66.103001}
\begin{eqnarray}
I_D(x_0)&=&-a\log\left(\dfrac{x_0}{x_m}-1\right)+b_D+\mathcal{O}(x_0-x_m),\\
I_R(x_0)&=&I_R(x_m)+\mathcal{O}(x_0-x_m).
\end{eqnarray}
where the coefficients in $I_D(x_0)$ are
\begin{eqnarray}
	a&=&{R(0,x_m)\over\sqrt{\hat{\beta}_m}},\\
b_D&=&{R(0,x_m)\over\sqrt{\hat{\beta}_m}}\ln{2(1-y_m)\over{A_m'x_m}},\\
\hat{\beta}_m&=&{C_m(1-y_m)^2 \left(y_m C_m''-C_m A_m''\right)\over{2y_m^2C_m'^2}}.
\end{eqnarray}

The deflection angle in the SDL becomes \cite{Bozza2002PRD66.103001}
\begin{equation}
\alpha(x_0)=-a\ln{\left({x_0\over{x_m}}-1\right)}+b+\mathcal{O}(x_0-x_m),
\end{equation}
where
\begin{equation}
  \label{}
  b=-\pi+b_D+b_R,
\end{equation}
and
\begin{equation}
  \label{}
  b_R=I_R(x_m).
\end{equation}
Since what we actually concern is the angular separation $\theta=u/D_{\mathrm{OL}}$ instead of $x_0$, where $D_{\mathrm{OL}}$ is the distance between the lens and the observer and the impact parameter $u$ is given by \cite{Weinberg1972Book,Virbhadra2000PRD62.084003}
\begin{equation}
u=\sqrt{\dfrac{C_0}{y_0}},
\end{equation}
the approximated relationship between $x$ and $u$ can be found as \cite{Bozza2002PRD66.103001}
\begin{equation}
  \label{}
  u-u_m = \hat{c}(x_0-x_m)^2
\end{equation}
where
\begin{equation}
	\hat{c} = \hat{\beta}_m\sqrt{\dfrac{y_m}{C_m^3}}\dfrac{{C_m'}{}^2}{2(1-y_m)^2}.
\end{equation}
For the charged Galileon black hole, it can be obtained that
\begin{equation}
  \label{}
  u_m=\dfrac{-16 \Gamma +3 \sqrt{9-32 \Gamma }+9}{2 \sqrt{2} \sqrt{-8 \Gamma +\sqrt{9-32 \Gamma }+3}}.
\end{equation}

Finally the deflection angle can be expressed by $\theta$ as \cite{Bozza2002PRD66.103001}
\begin{equation}
\alpha(\theta)=-\bar{a}\ln{\left({\theta D_{OL}\over{u_m}}-1\right)}+\bar{b}+\mathcal{O}(\theta D_{OL}-u_m),
\end{equation}
where the SDL coefficients are
\begin{eqnarray}
\bar{a}&=&{a\over2},\\
\bar{b}&=&-\pi+b_R+\bar{a}\ln{2\hat{\beta}_m\over{y_m}}.
\end{eqnarray}

For the charged Galileon black hole, we can find that
\begin{equation}
b_R=b_{R,0}+b_{R,1}\Gamma+\mathcal{O}(\Gamma^2),
\end{equation}
where
\begin{eqnarray}
b_{R,0}&=&\log (36)-4 \tanh ^{-1}\left(\frac{1}{\sqrt{3}}\right)=0.9496,\\
b_{R,1}&=&\dfrac{4}{9} \left[\frac{4}{\sqrt{3}}-5+\log \left(-6 \sqrt{3}+12\right)\right]=-0.9848.
\end{eqnarray}
And the SDL coefficients are
\begin{eqnarray}
\bar{a}&=&\frac{x_m^2 \sqrt{x_m-2 \Gamma }}{\sqrt{\left(\Gamma +x_m^2\right) \left(8 \Gamma ^2-9 \Gamma  x_m+\left(3-x_m\right) x_m^2\right)}},\\[1.2em]
\bar{b}&=&-\pi +\log (6)+\log (36)-4 \tanh ^{-1}\left(\frac{1}{\sqrt{3}}\right)\nonumber\\
&&+\left\{\dfrac{2}{9} \left[4+\log (6)\right]+\frac{4}{27} \left[4 \sqrt{3}-15+3 \log \left(-6 \sqrt{3}+12\right)\right]\right\}{\Gamma}+\mathcal{O}(\Gamma^2)\nonumber,\\
&=&-0.4002 + 0.3023 {\Gamma}+\mathcal{O}(\Gamma^2).
\end{eqnarray}

%%%%%%%%%%%%%%%%%%%%%%%%%%%%%%%%%%%%%%%%

\subsection{Calculating the time delay in the SDL}
\label{SecA.2}

For the time component of a null geodesic in the spacetime of eq. \eqref{appeq:ds2}, $\mathrm{d}t/\mathrm{d}x$ is given by \cite{Bozza2004GRG36.435}
\begin{equation}
	\dfrac{\mathrm{d}t}{\mathrm{d}x}=\tilde{P}_1(x,x_0)P_2(x,x_0),
\end{equation}
where
\begin{equation}
  \label{}
  	\tilde{P}_1(x,x_0)=\sqrt{\frac{B(x)A_0}{A(x)}},
\end{equation}
and $P_2(x,x_0)$ can be found in eq.~\eqref{eq:P2}.  The time taken by a photon from the source to the observer can be decomposed into three parts \cite{Bozza2004GRG36.435}
\begin{equation}
  \label{appeq:T3parts}
  T =  \tilde{T}(x_0) - \int^{\infty}_{D_{\mathrm{OL}}}\bigg|\frac{\mathrm{d}t}{\mathrm{d}x}\bigg|\mathrm{d}x - \int^{\infty}_{D_{\mathrm{LS}}}\bigg|\frac{\mathrm{d}t}{\mathrm{d}x}\bigg|\mathrm{d}x,
\end{equation}
where $\tilde{T}(x_0)$ is defined as \cite{Bozza2004GRG36.435}
\begin{equation}
\label{eq:dftime}
\tilde{T}(x_0)=\int^{\infty}_{x_0}\bigg|\dfrac{\mathrm{d}t}{\mathrm{d}x}\bigg|\mathrm{d}x.
\end{equation}
It is assumed that the observer and the source are far from the lens, the time delay between two relativistic images 1 and 2 can be given as \cite{Bozza2004GRG36.435}
\begin{eqnarray}
	 T_1-T_2&=&2\int^\infty_{x_{0,1}}\left|\dfrac{\mathrm{d}t}{\mathrm{d}x}(x,x_{0,1})\right|dx-2\int^\infty_{x_{0,2}}\left|\dfrac{\mathrm{d}t}{\mathrm{d}x}(x,x_{0,2})\right|\mathrm{d}x,\\
&=&\tilde{T}(x_{0,1})-\tilde{T}(x_{0,2})+2\int^{x_{0,2}}_{x_{0,1}}\dfrac{\tilde{P}_1(x,x_{0,1})}{\sqrt{A_{0,1}}}dx,
\end{eqnarray}
where the subscript ``${,i}$'' ($i=1,2$) of a quantity is the quantity of the $i$th relativistic image. With the same technique applied in the previous subsection of the appendix, the integral of $\tilde{T}(x_0)$ can be rewritten as \cite{Bozza2004GRG36.435}
\begin{equation}
  \label{}
  \tilde{T}(x_{0})=\int^1_0\tilde{R}(z,x_{0})f(z,x_{0})dz,
\end{equation}
where $f(z,x_0)=P_2(x,x_0)$ and
\begin{equation}
\tilde{R}(z,x_{0})=2\dfrac{1-y_{0}}{A'(x)}\tilde{P}_1(x,x_{0})\left(1-\dfrac{1}{\sqrt{A_{0}}f(z,x_{0})}\right).
\end{equation}
We can obtain its value in the SDL at $x_0\sim x_m$ and transform the variable $x_0$ to $u$. Finally, we can have \cite{Bozza2004GRG36.435}
\begin{eqnarray}
\tilde{T}(\theta)&=&-\tilde{a}\ln{\left(\dfrac{\theta D_{OL}}{u_m}-1\right)}+\tilde{b}+\mathcal{O}(\theta D_{OL}-u_m),\\
\tilde{a}&=&\dfrac{\tilde{R}(0,x_{m})}{2\sqrt{\hat{\beta}_m}},\\
\tilde{b}&=&b_D+b_R+\tilde{a}\log{\dfrac{\hat{c}x_m^2}{u_m}}.
\end{eqnarray}

Here we assume the source, the lens and the observer are aligned almost in a line. By using eqs. \eqref{eq:theta0} and \eqref{eq:theta1} and an approximated relation that \cite{Bozza2004GRG36.435}
\begin{equation}
	 \int^{x_{0,2}}_{x_{0,1}}\dfrac{\tilde{P}_1(x,x_{0,1})}{\sqrt{A_{0,1}}}\mathrm{d}x\approx\sqrt{\dfrac{B_m}{A_m}}\ (x_{0,2}-x_{0,1}),
\end{equation}
 we can have the time delay between a $n$-loop and a $m$-loop relativistic image as \cite{Bozza2004GRG36.435}
\begin{equation}
\label{}
\Delta T_{n,m} = \Delta T_{n,m}^{\ 0}+\Delta T_{n,m}^{\ 1},
\end{equation}
where
\begin{eqnarray}
\Delta T_{n,m}^{\ 0}&=&2 \pi (n-m) u_m,\\[0.5em]
\Delta T_{n,m}^{\ 1}&=&2\sqrt{\frac{B_m}{A_m}} \sqrt{\dfrac{u_m}{\hat c}}\exp\left({\frac{\bar{b}}{2\bar{a}}}\right)\left[\exp\left({-\frac{m\pi}{\bar{a}}}\right)-\exp\left({-\frac{n\pi}{\bar{a}}}\right)\right].
\end{eqnarray}

%%%%%%%%%%%%%%%%%%%%%%%%%%%%%%%%%%%%%%%%%%%%%%%%%

\acknowledgments

This work is funded by the National Natural Science Foundation of China (Grant No. 11573015) and the Opening Project of Shanghai Key Laboratory of Space Navigation and Position Techniques (Grant No. 14DZ2276100).

% The bibliography will probably be heavily edited during typesetting.
% We'll parse it and, using the arxiv number or the journal data, will
% query inspire, trying to verify the data (this will probalby spot
% eventual typos) and retrive the document DOI and eventual errata.
% We however suggest to always provide author, title and journal data:
% in short all the informations that clearly identify a document.
\bibliographystyle{JHEP}
\bibliography{Refs20160620}

% Please avoid comments such as "For a review'', "For some examples",
% "and references therein" or move them in the text. In general,
% please leave only references in the bibliography and move all
% accessory text in footnotes.

% Also, please have only one work for each \bibitem.

\end{document}